\documentclass[11pt, preprint]{aastex}
\usepackage[utf8]{inputenc}
\usepackage{graphicx}
\usepackage{latexsym}
\usepackage{amsfonts,amsmath,amssymb, gensymb}
\usepackage{scrextend}  
\usepackage{hhline}
\usepackage{booktabs}
\usepackage{multirow}
\usepackage{pdflscape} 
\usepackage{float} 
\usepackage{geometry} 
\usepackage{adjustbox}
\usepackage{natbib}
\usepackage{pbox}

\usepackage[left]{lineno}
\usepackage{color}

\usepackage{url}
\usepackage{color}
\usepackage{hyperref}
\hypersetup{
    colorlinks=true,
    linkcolor=black,
    filecolor=black,
    citecolor=black,      
    urlcolor=cyan}
\usepackage{soul}		

\begin{document}
\title{The second catalog of flaring gamma-ray sources from the Fermi All-sky Variability Analysis}
\author{
S.~Abdollahi\altaffilmark{1}, 
M.~Ackermann\altaffilmark{2}, 
M.~Ajello\altaffilmark{3,4}, 
A.~Albert\altaffilmark{5}, 
L.~Baldini\altaffilmark{6}, 
J.~Ballet\altaffilmark{7}, 
G.~Barbiellini\altaffilmark{8,9}, 
D.~Bastieri\altaffilmark{10,11}, 
J.~Becerra~Gonzalez\altaffilmark{12,13}, 
R.~Bellazzini\altaffilmark{14}
% ...
}
\altaffiltext{1}{Department of Physical Sciences, Hiroshima University, Higashi-Hiroshima, Hiroshima 739-8526, Japan}
\altaffiltext{2}{Deutsches Elektronen Synchrotron DESY, D-15738 Zeuthen, Germany}
\altaffiltext{3}{Department of Physics and Astronomy, Clemson University, Kinard Lab of Physics, Clemson, SC 29634-0978, USA}
\altaffiltext{4}{email: majello@slac.stanford.edu}
\altaffiltext{5}{Los Alamos National Laboratory, Los Alamos, NM 87545, USA}
\altaffiltext{6}{Universit\`a di Pisa and Istituto Nazionale di Fisica Nucleare, Sezione di Pisa I-56127 Pisa, Italy}
\altaffiltext{7}{Laboratoire AIM, CEA-IRFU/CNRS/Universit\'e Paris Diderot, Service d'Astrophysique, CEA Saclay, F-91191 Gif sur Yvette, France}
\altaffiltext{8}{Istituto Nazionale di Fisica Nucleare, Sezione di Trieste, I-34127 Trieste, Italy}
\altaffiltext{9}{Dipartimento di Fisica, Universit\`a di Trieste, I-34127 Trieste, Italy}
\altaffiltext{10}{Istituto Nazionale di Fisica Nucleare, Sezione di Padova, I-35131 Padova, Italy}
\altaffiltext{11}{Dipartimento di Fisica e Astronomia ``G. Galilei'', Universit\`a di Padova, I-35131 Padova, Italy}
\altaffiltext{12}{NASA Goddard Space Flight Center, Greenbelt, MD 20771, USA}
\altaffiltext{13}{Department of Physics and Department of Astronomy, University of Maryland, College Park, MD 20742, USA}
\altaffiltext{14}{Istituto Nazionale di Fisica Nucleare, Sezione di Pisa, I-56127 Pisa, Italy}
% ...
\altaffiltext{4}{email: matteo.giomi@desy.de}


\begin{abstract}
We present the second catalog of flaring gamma-ray sources (2FAV) detected with the Fermi All-sky Variability Analysis (FAVA), a tool that blindly searches for transients over the entire sky observed by the Large Area Telescope (LAT) on board the \textit{Fermi} Gamma-ray Space Telescope. With respect to the first FAVA catalog, this catalog benefits from a larger data set, the latest LAT data release (Pass 8), as well as from an improved analysis that includes likelihood techniques for a more precise localization of the transients.
Applying this analysis on the first 7.4 years of \textit{Fermi} observations, and in two separate energy bands 0.1$-$0.8 GeV and 0.8$-$300 GeV, a total of 4547 flares has been detected with a significance greater than $6\sigma$ (before trials), on the time scale of one week. Through spatial clustering of these flares, 518 variable gamma-ray sources are identified. Likely counterparts, based on positional coincidence, have been found for 441 sources, mostly among the blazar class of active galactic nuclei. For 77 2FAV sources, no likely gamma-ray counterpart has been found. For each source in the catalog, we provide the time, location, and spectrum of each flaring episode.
Studying the spectra of the flares, we observe a harder-when-brighter behavior for flares associated with blazars, with the exception of BL Lac flares detected in the low-energy band. The photon indexes of the flares are never significantly smaller than 1.5. For a leptonic model, and under the assumption of isotropy, this limit suggests that the spectrum of the freshly accelerated electrons is never harder than $p\sim$2.
\end{abstract}

\keywords{Fermi LAT, variability, gamma-rays, all-sky monitoring, flares}

\section{Introduction}
The Large Area Telescope \citep[LAT,][]{Atwood_2009} on board the \textit{Fermi} Gamma-ray Space Telescope observes $\sim$20\% of the sky at any given moment. It spends $\sim$80\% of the time in survey mode, imaging the entire sky roughly every three hours. This, together with its wide energy range, high angular resolution, and low detection threshold, makes the LAT well suited to investigate variable and transient phenomena in the gamma-ray sky.


Various analysis pipelines are maintained by the \textit{Fermi}-LAT Collaboration to search for and monitor gamma-ray transients. Monthly light curves are reported in all \textit{Fermi}-LAT catalogs, e.g in the third \textit{Fermi}-LAT source catalog \cite[3FGL,][]{Acero_2015}. The Monitored Source List\footnote{\href{http://fermi.gsfc.nasa.gov/ssc/data/access/lat/msl\_lc/}{http://fermi.gsfc.nasa.gov/ssc/data/access/lat/msl\_lc/}} provides daily and weekly light curves of the brightest sources and transients found during LAT observations. Variability on time scales of 6 hours to 1 day is monitored by the \textit{Fermi} Flare Advocate program~\citep{Ciprini_2012} using quick-look science data products of the Automated Science Processing pipeline ~\citep{ASP}. Finally, the \textit{Fermi} All-sky Variability Analysis \citep[FAVA, ][]{Ackermann_2013} uses photometric analysis to systematically search for transients over the entire sky. FAVA complements the Flare Advocate variability search: it uses a different technique to detect the transients (the Flare Advocates variability search is based on wavelet decomposition of the all-sky counts maps).

FAVA searches for transient gamma-ray emission by comparing, for every direction in the sky, the number of counts observed in a given time interval, to the expected number of counts, as derived from a long-term average. With respect to maximum likelihood analysis methods~\citep[see, e.g.,][]{Mattox}, FAVA has several advantages. It is independent of any model of the diffuse gamma-ray emission. This emission is expected to be constant over time scales comparable to the duration of the \textit{Fermi} mission. It will therefore cancel out when comparing the observed to the expected counts. FAVA also does not rely on any assumption on the spectrum of the source, and it is sensitive to both positive and negative flux variations alike. FAVA is also less computationally intensive when compared to maximum likelihood analysis. It enables an uninformed search for flux variations over the entire sky in a variety of different energy bands and timescales. FAVA was used to build a first catalog of variable LAT sources \cite[1FAV,][]{Ackermann_2013}. FAVA has been used to continuously monitor the sky on weekly time bins, its fully automated analysis complementing the information available to the \textit{Fermi} Flare Advocates \citep{MarcoAtel, DanAtel}.

Previous to this work, the major limitation of FAVA was its poor localization precision, especially at low energies. The 68\% containment radius of the LAT point-spread function (PSF) is $0.8\degree$ at 1 GeV and increases with decreasing energy, reaching $\sim 5\degree$ at 100 MeV\footnote{\label{irfs}The instrument response functions for the Pass 8 event reconstruction can be found at: \href{https://www.slac.stanford.edu/exp/glast/groups/canda/lat\_Performance.htm}{https://www.slac.stanford.edu/exp/glast/groups/canda/lat\_Performance.htm}}. Low-energy flares, as localized by FAVA, are often scattered over the scale of one degree or more, even if coming from the same astrophysical source. To address this issue, FAVA has been upgraded to include an automated follow-up analysis with a maximum-likelihood detection and localization method. The localization precision has greatly improved with respect to the previous version of FAVA. With the maximum likelihood analysis, the energy spectrum of each flare is also measured, providing better characterization of the transient source. With respect to the 1FAV catalog, this upgraded FAVA analysis also benefits from the increased sensitivity of the latest LAT data release \citep[Pass 8,][]{pass8}.

Here we describe the new version of the FAVA analysis, and present a list of 518 flaring gamma-ray sources found by applying it to the first 7.4 years of LAT observations. For each source in this catalog, likely gamma-ray counterparts, and detailed information on every flare are provided. The FAVA analysis pipeline described here will also continue to monitor the gamma-ray sky, searching for new transients. The results of this on-line analysis, as well as photometric aperture light curves for an all-sky grid of coordinates are made publicly available at NASA’s \textit{Fermi} Science Support Center\footnote{\href{http://fermi.gsfc.nasa.gov/ssc/data/access/lat/FAVA/}{http://fermi.gsfc.nasa.gov/ssc/data/access/lat/FAVA/}}.

\section{FAVA analysis}\label{sec:Analysis}
As for the previous catalog, FAVA uses weekly time bins. Two independent energy bands are used: 0.1$-$0.8 GeV and 0.8$-$300 GeV to enhance the sensitivity to spectrally soft and hard flares, respectively. The data used belong to the \texttt{P8R2\_SOURCE} event class with an additional cut on the zenith angle at 95$\degree$ to limit the contamination from the Earth limb.

The FAVA analysis now consists of two sequential steps. The first one comprises all the photometric analysis that was used to produce the 1FAV. This step, which we will call the photometric FAVA analysis, is now used to provide a list of seed flares that are further analyzed, in the second step, with likelihood techniques.

The photometric FAVA analysis is described in detail in~\cite{Ackermann_2013}; in this paragraph we recall its main steps for convenience of the reader. For every time bin, all-sky maps of the number of observed and expected events are created, with a resolution of 0.25 square degrees ($0.5\degree\times 0.5\degree$) per pixel. 

\begin{landscape}
\begin{figure}[h!]
\centering										
\includegraphics[width=1\columnwidth]{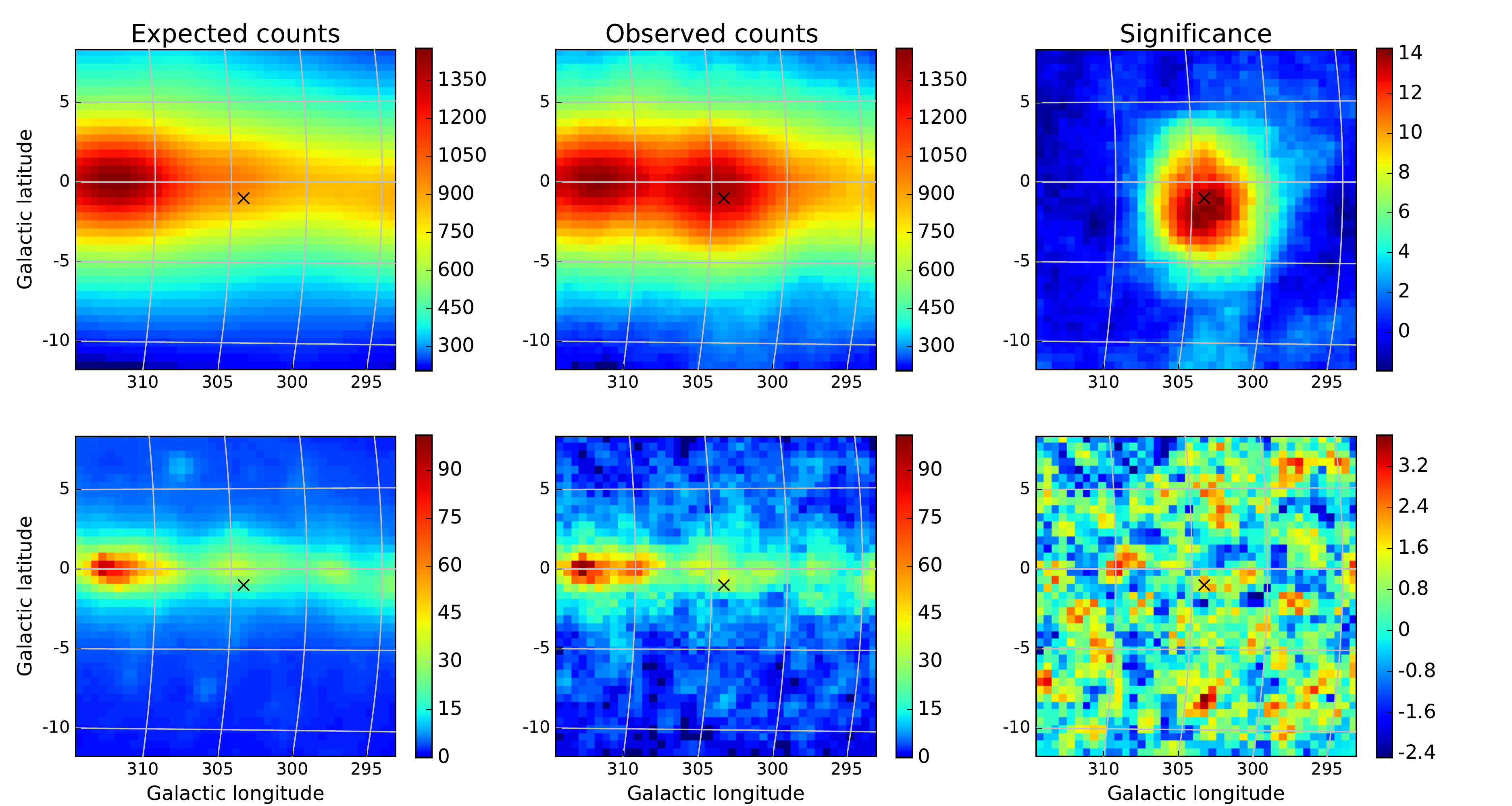}
\caption{
Examples of FAVA maps for the bright flare associated with one of the periastron passages of PSR B1259$-$63/LS 2883. The maps represent the analysis for the time bin [316971818, 317576618] MET, around MJD 55582. Only a $20\degree \times 20\degree$ region centered on the flare position is shown. The maps are arranged as follows: expected counts (left), observed counts (center), significance expressed in Gaussian sigma (right). Upper row: 0.1$-$0.8 GeV, lower row: 0.8$-$300 GeV. The black \lq x\rq~marks the position of PSR B1259$-$63.}
\label{fig:favamaps_examples}
\end{figure}
\end{landscape}

The number of expected counts in each pixel and time bin is derived from the total number of events recorded from that pixel during the first four years of the mission, after properly weighting for the different exposures. Four years, the time covered by the 3FGL, is much longer than the one-week duration of the time bins and the statistical uncertainty on the number of expected counts is therefore negligible. This time-averaged emission is then used over the full 7.4 year time span of the 2FAV. Both the expected and observed counts maps are smoothed to account for the finite size of the LAT PSF. The smoothing assigns to each pixel all events that are within a distance corresponding to the 68\% containment radius of the PSF. As the PSF depends on the energy of the photons and on its incidence angle with respect to the LAT, we integrate over these two parameters. With the numbers of observed and expected counts, the amplitude of flux variations is converted into a probability using Poisson statistics, which is then translated into Gaussian significance for convenience. To enhance the sensitivity for spectrally hard and soft flares, the analysis is performed separately at high and low energy. For every time bin, FAVA produces all-sky maps of the significance of the observed flux variations in the two energy bands. Examples of these significance maps, as well as of the expected and observed counts maps, are presented in Figure~\ref{fig:favamaps_examples}. Flares are identified as local maxima, and minima in case of negative flux variations, in the significance maps. This is accomplished using a peak-finder algorithm~\citep{peakfinder} that is applied to both the low- and high-energy significance maps. To take advantage of the higher positional precision of the high-energy analysis, low-energy flares are merged to the high-energy ones found in the same week if within 3$\degree$ and if the high-energy flare has a significance greater than $5.5\sigma$. As already found in the 1FAV, the value of 3$\degree$ gives a good trade-off between the density of flares detected in one week and the positional accuracy of the peak finder.

The time interval used to generate the expected long-term counts maps has a duration of 4 years and is fixed to the interval used to produce the 3FGL catalog. As the statistical errors on FAVA are dominated by the low-counting statistics of the weekly exposures, this choice introduces a comparatively negligible error compared to the choice of a longer integration time for the long-term counts maps. The significance of the flares measured by FAVA is therefore referred to as an excess of flux over a static 3FGL sky. As we shall see, this will allow a meaningful comparison between FAVA significances and TS values. In creating these long-term counts maps, a circular region of $10\degree$ radius around the position occupied by the Sun on 2012 Mar 7 has been masked for the corresponding time bin, to account for the bright solar flare that occurred on that date~\citep{Ajello_2014}. If not corrected, this solar flare would increase the number of expected counts around its position, giving rise to spurious negative flares. Other solar flares that occurred in the 3FGL time range have no measurable effect on the long-term counts maps.

The likelihood follow-up analysis is based on the generation of test-statistic (TS) maps using the \texttt{gttsmap} tool from the \textit{Fermi} Science Tools (v10r01p00, internal to Fermi LAT collaboration). The TS represents twice the logarithm of the ratio of the likelihood evaluated at the best-fit parameters when including a candidate point source with a power-law spectrum (free in index and normalization), to the likelihood under a background-only, null hypothesis. The TS maps are the result of a series of such likelihood ratio tests, performed on a grid of locations in the sky. The TS maps are centered at the flare position measured by the peak-finder algorithm. To limit the amount of computation, the likelihood follow-up is triggered only by flares for which FAVA found a significance greater than $4\sigma$ in at least one of the two energy bands\footnote{The likelihood analysis tools we are using cannot test for an absence of flux. Negative flares will only be measured by the photometric FAVA analysis.}. We will refer to this condition simply as the "seed condition". If the seed condition is satisfied, the TS maps are generated for both energy ranges. 

\begin{figure}[h!]
    \begin{minipage}[b]{0.5\columnwidth}
    \centering
    \includegraphics[width=\columnwidth]{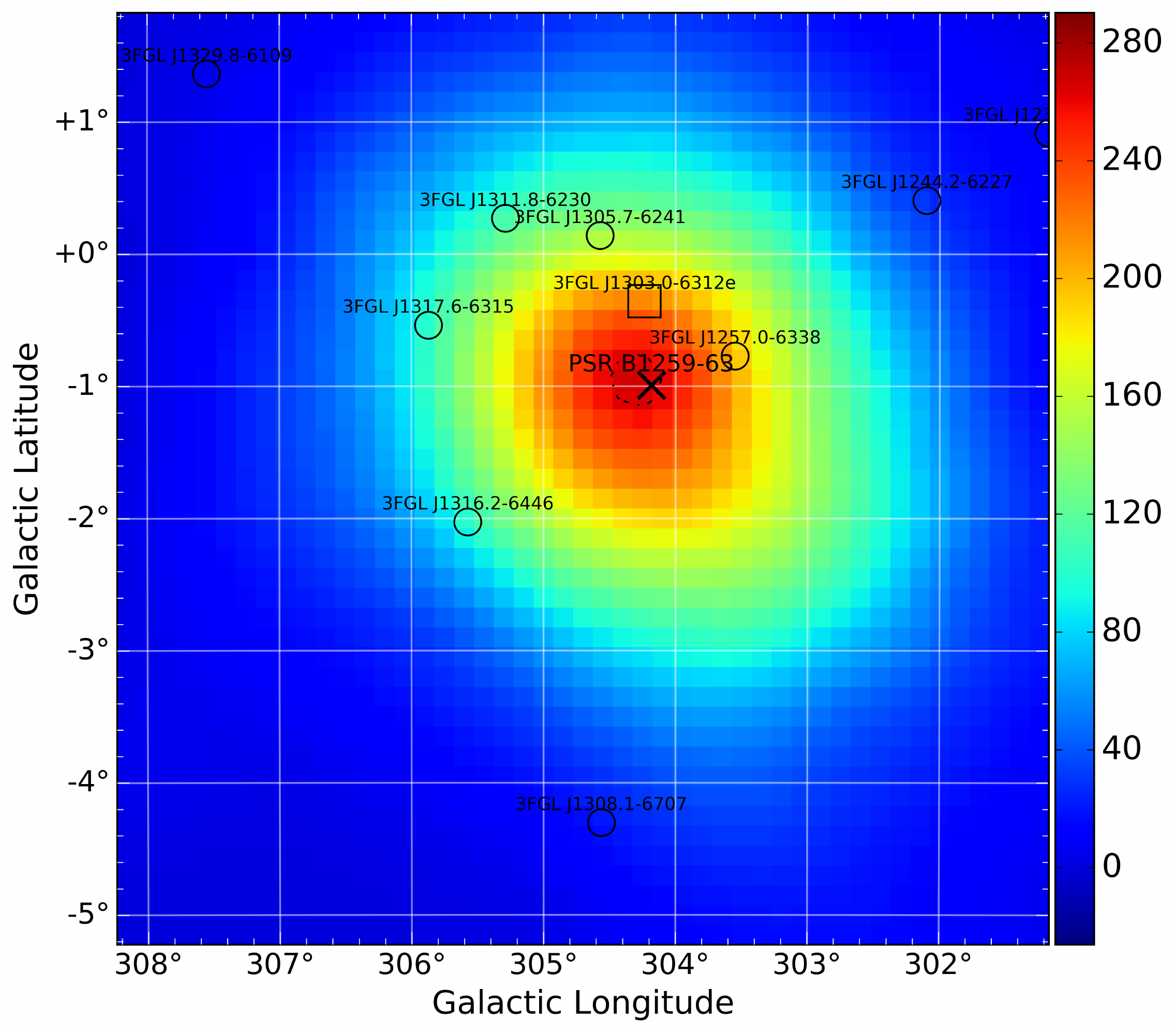}
    \end{minipage}
    \hfill
    \begin{minipage}[b]{0.5\columnwidth}
    \centering
    \includegraphics[width=\columnwidth]{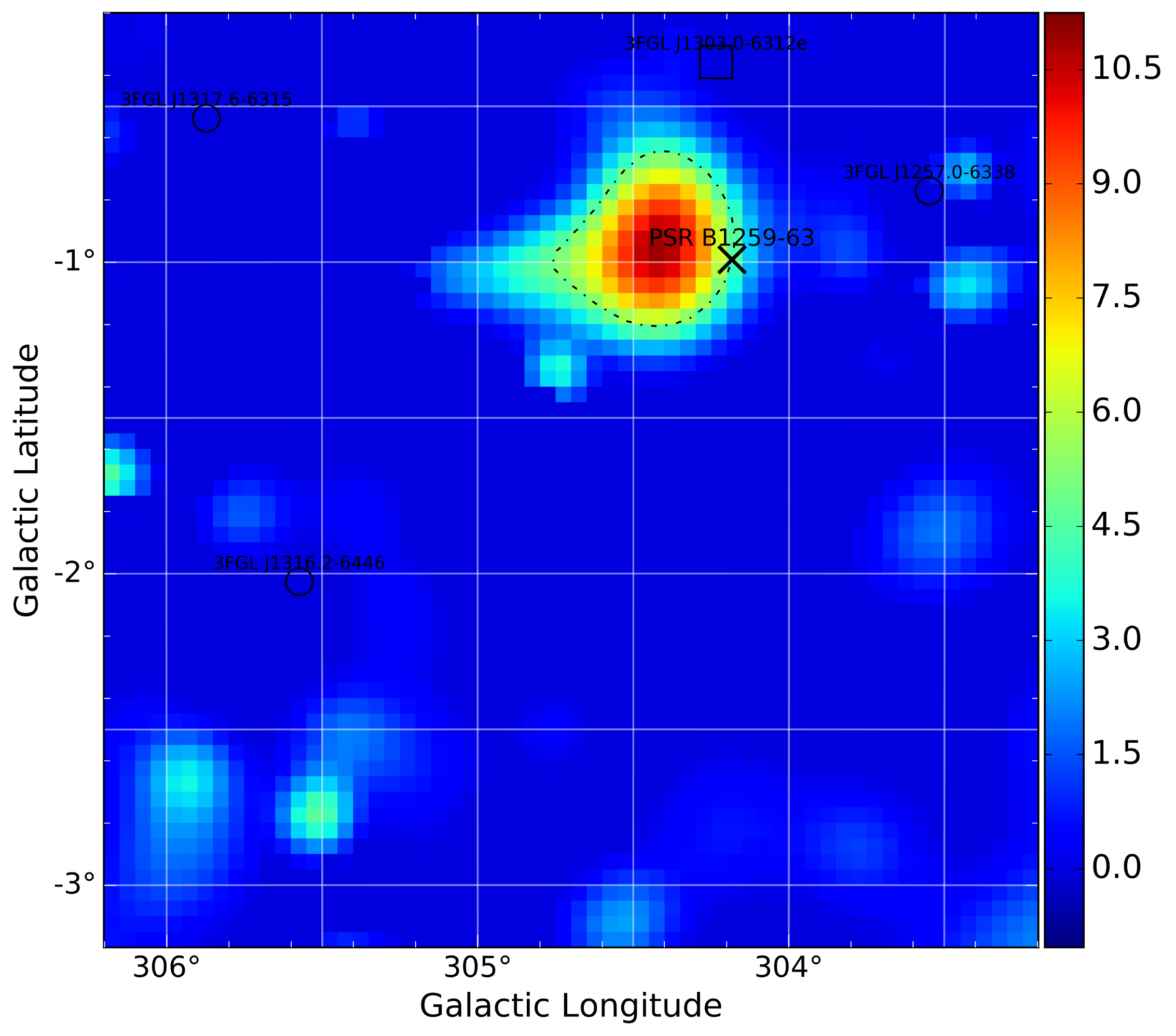}
    \end{minipage}
    \caption{Examples of the TS maps generated by the automatic likelihood follow-up analysis. The maps are referred to the bright low-energy flare associated with the first periastron passage of the binary system PSR B1259$-$63/LS 2883 detected by FAVA; see Figs.~\ref{fig:favamaps_examples} and \ref{fig:lc_PSRB125963}. The values of TS can be read from the color scale. The left panel shows the TS map at low energies, 0.1$-$0.8 GeV. The right panel shows the TS map at high energies, 0.8$-$300 GeV. Note the different angular sizes ($7\degree \times 7\degree$ for the TS map at low energies and $3\degree \times 3\degree$ at high energy) and centering of the maps. The dashed line around the maximum marks the 95\% CL contour. This flare is soft ($\Gamma = 3$) resulting in a much higher detection significance in the low-energy band. The positions and uncertainties of the closest 3FGL sources included in the sky model used for the analysis are also shown as circles (or squares, in case of extended sources). The black X marks the position of PSR B1259$-$63.}
    \label{fig:tsmaps_examples}
\end{figure}

In the low-energy band, the likelihood analysis is binned in 10 logarithmic energy bins between 100 and 800 MeV and the TS maps have a size of $7\times7$ deg$^2$ with a pixel size of $0.15\degree \times 0.15\degree$. Unbinned analysis is performed in the high-energy band, and the TS maps have a $3\degree \times 3\degree$ size and a pixel size of $0.05\degree \times 0.05\degree$. The size and resolution of the TS maps are chosen as a compromise between localization accuracy, tolerance towards misplaced position of the seed, and the requirement for the likelihood analysis to complete within less than a day. For both energy bands, the model used to derive the TS includes the point sources in the region of interest ($15\degree$ and $8\degree$ in radius at low and high energy respectively), and the templates for Galactic and isotropic diffuse emission recommended for the considered event class. The point sources, and their energy spectra used in the model, are taken from the 3FGL. In the fit, the index and normalization of the Galactic diffuse background are free parameters, as is the normalization of the diffuse isotropic emission. All the point and extended sources are instead fixed to their 3FGL values. The resulting values of TS are therefore referred to an eventual excess in flux, above a static 3FGL source population. Examples of the TS maps are presented in Figure~\ref{fig:tsmaps_examples}.

From the TS maps, the position of the flare is measured as the center of the pixel that has the highest TS. The 95\% error radius of the flare position, $r_{95}$, is estimated as the average distance between the maximum of the TS map and its contour at the 95\% confidence level (CL), corresponding to a $\Delta TS=5.99$. The $r_{95}$ is not allowed to be smaller than the size of the TS map pixel. The shape of the 95\% CL contour and its distance from the border of the map are also analyzed to measure the reliability of the flare localization, see the next section for details. The flux and spectral index of the flare are then measured with an additional likelihood fit (using \texttt{gtlike}). In this fit the flare is modeled as a point source placed at the position of the maximum TS and  whose power-law spectrum is allowed to vary in index and normalization. This analysis step shares the same analysis settings as the TS map generation.

\begin{figure}[h!]
\begin{center}
\includegraphics[width=0.7\columnwidth]{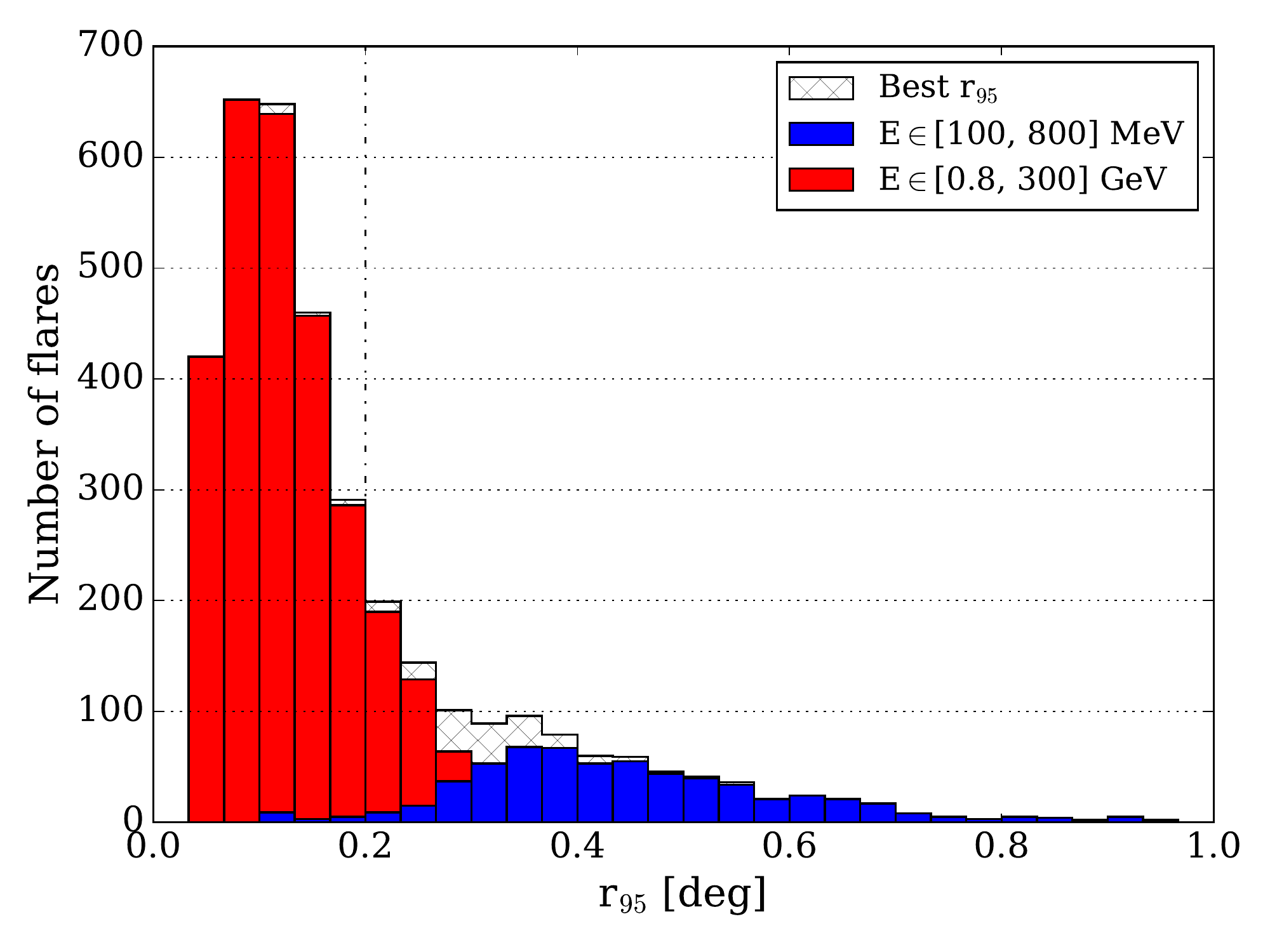}
\caption{\label{fig:best_r95}
Distribution of the $r_{95}$ for the individual 2FAV flares. Red: positions from high-energy analysis; blue: positions from low-energy analysis; hatched black: $r_{95}$ from the analysis that provides the best flare position. The dotted line marks the division between the better localized flares used in the very first step of the clustering, and the ones added in the second step. See Section~\ref{sec:build_catalog} for details.}
\end{center}
\end{figure}

For every flare satisfying the seed condition, there are three different estimators of its position: the merged position from the peak-finder algorithm, and the more refined localizations from the TS maps at low and high energies. As the peak-finder algorithm does not provide an error on the peak position, its accuracy can be estimated comparing the flare position to known, bright flaring gamma-ray sources, as described in~\cite{Ackermann_2013}. The resulting average values for the $r_{95}$ are $1\degree$ and $0.8\degree$ for low- and high-energy flares respectively. For comparison, Figure~\ref{fig:best_r95} shows the distribution of $r_{95}$ from TS map positions of catalog flares. These distributions peaks at $0.35\degree$ at low energies and $0.1\degree$ at high energies, an improvement of a factor of $\sim$3 and $\sim$8 respectively as compared to the photometric FAVA analysis. These improvements depend on how the peak finder and the likelihood analysis respond to the steep decrease of the PSF radius with increasing energies. Although in general high-energy TS maps provide more accurate flare localization, there are cases when the low-energy analysis provides more accurate positions. The best estimator of the flare position is chosen comparing the corresponding values of $r_{95}$. 

Beside a better positional accuracy, the likelihood analysis follow-up has introduced other advantages. For each flare, a measurement of the uncertainty in the localization is available, facilitating searches for counterparts of the single flares. The likelihood analysis also has better sensitivity. Converting the value of TS into equivalent significance $S_{TS}$ (assuming 2 degrees of freedom) and comparing it with the FAVA significance $S_{FAVA}$, we found that, on average, $S_{TS}\approx1.26S_{FAVA}$ at low energies and $S_{TS}\approx1.1S_{FAVA}$ at high energies.

\section{Construction of the catalog}\label{sec:build_catalog}

To build the 2FAV catalog, we apply FAVA on the first 387 weeks of \textit{Fermi} observations, from Mission Elapsed Time (MET) 239557418 to 473615018, or Modified Julian Date (MJD) from 54682 (2008-08-04) to 57391 (2016-01-04). In this time range, a total of 7106 seed flares has been found, roughly 18 per week. 

To limit false flares due to statistical fluctuations, strict cuts are applied on the significance of the flares used to construct the catalog. Separate cuts are applied for the different energy bands, and for the likelihood and the photometric FAVA analysis. To be included in the catalog, flares must have:
\begin{itemize}
\item a significance greater than $6\sigma$ (or smaller than $-6\sigma$), or a TS greater than 39 in a single energy band, or 
\item a significance greater than $4\sigma$ (or smaller than $-4\sigma$), or a TS greater than 18 in both energy bands simultaneously and a distance between the low- and high-energy flare position smaller than 3 and 1.5 degrees for photometric and TS map positions, respectively. As the two energy bands are independent, this requirement yield a total significance of $\approx 6\sigma$.
\end{itemize}
The number of trials is estimated as for the 1FAV, counting the number of independent positions on the sky (ratio of $4\pi$ to the solid angle of the PSF), and multiplying it by the number of time bins. The expected number of false positives corresponding to this significance threshold is $\approx$0.001 at low energies and $\approx$0.03 at high energies. In each of these three energy ranges (high energy, low energy, and combined) we require the Sun to be more than 6$\degree$ away from the position of the flare. Flare information from the likelihood follow-up is not considered if the TS in both energy bands is below 18, or if the 95\% CL contour intersects the border of the TS map, or if it is composed of more than one closed path. When the TS-map contour has such features, we also discard positive flares detected by the photometric FAVA analysis if the corresponding TS is larger than 18, a situation that can arise from incorrect position of the FAVA flare. The number of flares in the various cut classes are summarized in Table~\ref{tab:flarecuttable}.

\begin{figure}[h!]
\centering
\includegraphics[width=1\columnwidth]{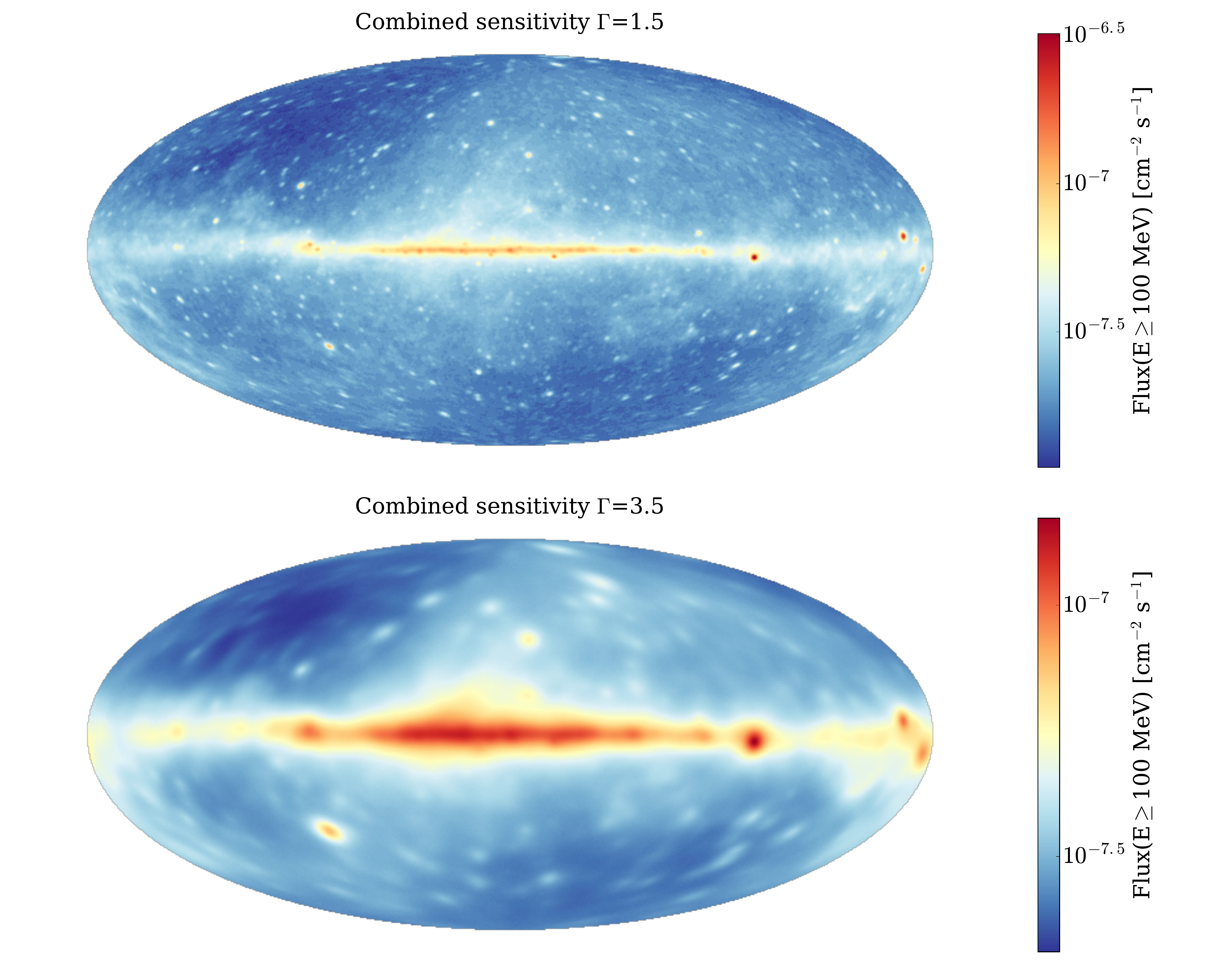}
\caption{
Sensitivity for flare detection of the photometric FAVA analysis (see Section~\ref{sec:build_catalog} for details). These maps shows the minimum flux increase, in a one week time bin, that corresponds to a FAVA significance greater than $6\sigma$ in one of the two FAVA energy bands or greater than $4\sigma$ in both energy bands simultaneously. The maps are computed assuming a power-law spectrum for the flaring source, the average weekly exposure, and for two reference values of the photon index $\Gamma$. Top panel: spectrally hard flares ($\Gamma=1.5$), bottom: soft flares ($\Gamma=3.5$). The maps are shown in Galactic coordinates and in Hammer-Aitoff projection.}
\label{fig:sensmaps}
\end{figure}

The sensitivity of the photometric FAVA analysis for a given significance threshold is determined by the maps of expected counts, assuming an average weekly exposure. The sensitivity maps presented in Figure~\ref{fig:sensmaps} show the minimum flare flux (above 100 MeV) needed to reach the FAVA detection threshold of either $6\sigma$ in one of the two FAVA energy bands, or $4\sigma$ in both energy bands simultaneously. These maps have been computed assuming a power-law spectrum for the flaring source, and for two reference values of the photon index $\Gamma$: 1.5 and 3.5. Most of the 2FAV flares, $\sim 90\%$ of the ones detected in the low-energy band and $\sim 96\%$ of the flares detected at high energies, have photon index between these two values (see Section~\ref{subsec:flarespec}). Harder flares ($\Gamma\lesssim2$) and softer flares ($\Gamma\gtrsim2.5$) are more likely to be detected in the high- or low-energy band, respectively. For intermediate values of $\Gamma$, requiring more than $4\sigma$ in both energy bands provides better sensitivity. These sensitivity maps do not account for the likelihood follow-up. They provide only an upper limit on the actual sensitivity of the 2FAV analysis. Lower flare fluxes, if bright enough to trigger the likelihood follow-up, could also be detected due to the slightly better sensitivity of the likelihood analysis, which yields roughly 26\% and 10\% higher significance than the photometric FAVA analysis at low and high energies, respectively. As a first approximation, these values can be used to scale the sensitivity fluxes.

\subsection{Clustering of the flares}\label{subsec:clustering}

The sources of the 2FAV catalog are identified through spatial clustering of the 4547 flares that satisfy the previously discussed cuts. As for the previous catalog, the clustering uses a Minimum Spanning Tree~\citep[MST,][]{mst} algorithm. To take full advantage of the increased accuracy of TS map positions, the clustering is performed in different steps, starting with the best localized flares and gradually including less-precise information.

The catalog flares are divided in three groups, according to the quality of their localization: group A consists of 2471 flares that have TS map position with $r_{95}\leq 0.2\degree$. Group B consists of 1087 flares whose position is also derived from TS maps, but whose $r_{95}>0.2\degree$. Group C consists of the 989 flares that have only photometric FAVA positions, 967 of which are negative. For the groups A and B, the $r_{95}$ provides a direct measurement of the error on the flare position. For the clustering, the separation between pairs of flares belonging to these two groups will be measured in units of $r_{68}$, dividing the angular separation by the combined error on their positions.

\begin{landscape}
\begin{figure}[h!]
\centering
\includegraphics[width=1\columnwidth]{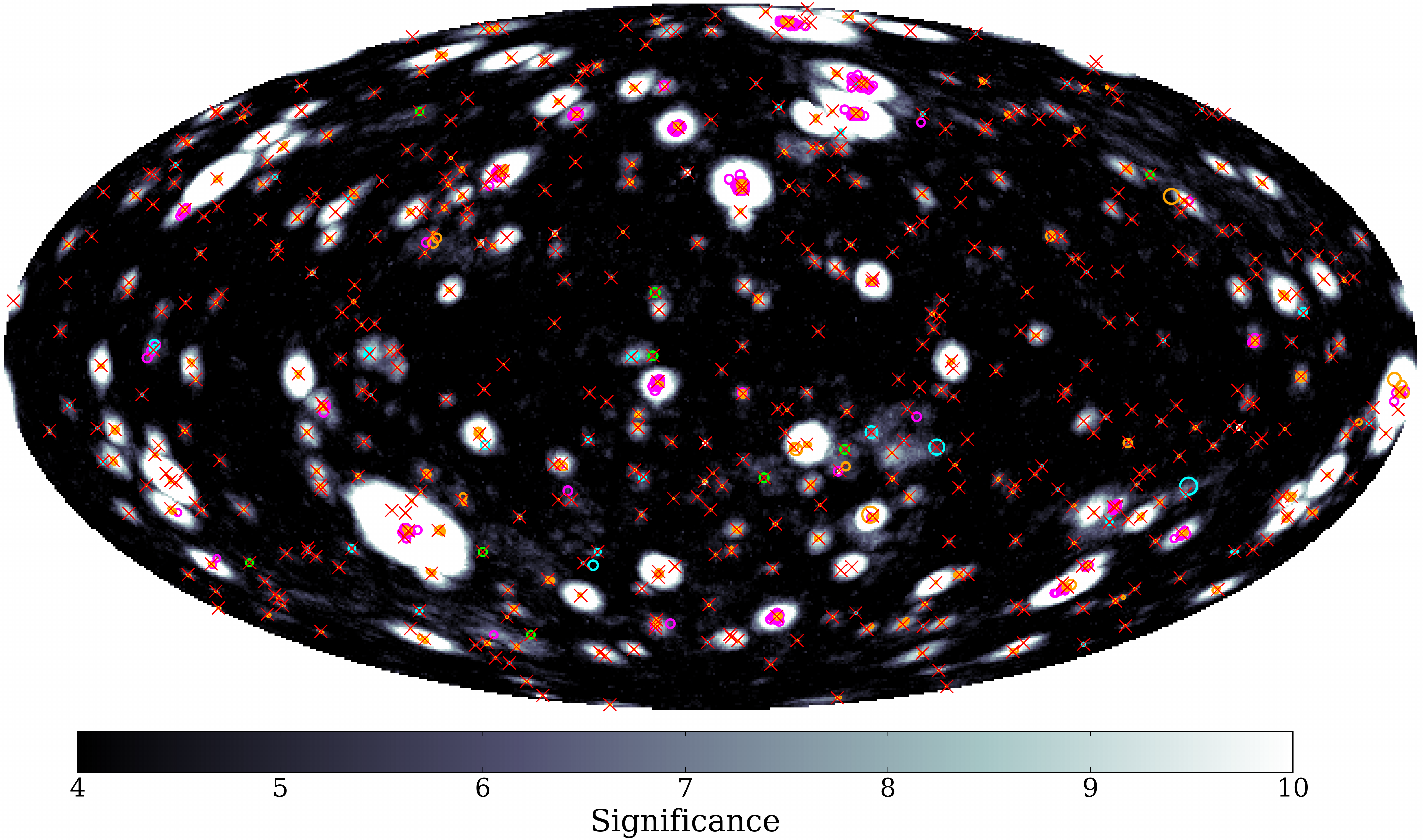}
\caption{Positions of 2FAV sources and flares on the sky, in Galactic coordinates and Hammer-Aitoff projection. The background image shows the maximum significance detected for each pixel, in either the low-, or high-energy band. The red crosses represent the 2FAV sources. The flares used to construct the 2FAV are also shown: yellow circles are flares with the best determined position from the TS maps. Flares with worse TS map positioning are in orange, if they have been assigned to a cluster, or cyan otherwise. Flares with only FAVA positions are in magenta if they have been assigned to a cluster, green if they constitute a separate cluster. For all the flares, the radius of the drawn circles is equal to $r_{95}$.}
\label{fig:catandflares}
\end{figure}
\end{landscape}

As a first step, we build the MST of the flares of group A. Clusters of flares are identified by cutting the graph edges longer than $4r_{68}$: neighboring flares are assigned to the same cluster if within $4r_{68}$ from one another\footnote{The corresponding cut of the scaled flare-flare distance $d'=d(i, j)[\text{deg}] / \sqrt{(r_{95}^2(i)+r_{95}^2(j)})$ is $d' \leq 2.5$. Since for a 2-dimensional symmetric Gaussian distribution $r_{68}\simeq0.617r_{95}$, this cut corresponds to $4r_{68}$.}. In this first step, 437 clusters are found. Next, we include group-B flares, the ones with worse TS map localization. Among these, 988 flares are assigned to clusters of group-A flares as the flare-cluster distance is smaller than $4r_{68}$. The remaining 99 group-B flares are independently clustered, again with a $4r_{68}$ cut on the graph edges, yielding 72 more clusters. Finally, group-C flares, the ones with only FAVA positions, are added. They are assigned to a pre-existing cluster if within $3.5\degree$ from it, otherwise independently clustered. In this way, 980 group-C flares are added to pre-existing clusters. The remaining 9 group-C flares are widely separated from each other and result in 9 additional clusters. Figure~\ref{fig:catandflares} presents the positions of the 2FAV flares, using different colors to differentiate between the groups defined here. The background image is the maximum photometric significance detected in each pixel in either the low- or high-energy band. The flares cluster on top of $>6\sigma$ maxima in the significance map. Thanks to the improved localization accuracy, the flares cluster on angular scales that are much smaller than the width of the significance maxima. 

The 2FAV sources are defined by these flare clusters. As each cluster can contain more than one flare, the positions of the flares in a cluster are combined to estimate the position of the source. The source position is computed as the weighted average of the flare positions, with weights given by $r_{95}^{-2}$. For flares without TS-map localization, the values of $r_{95}$ used as weights are taken to be equal to the average values of 0.8 and 1 degrees at high and low energy, respectively. The 95\% statistical error radius on the source position is computed as $1/\sqrt{\sum_{i} r_{95}^{-2}(i)}$. For each cluster, the source position and the associated statistical error are computed taking into account only the flares with the best available localization group, following the hierarchy outlined in the previous paragraph. Additionally, the error on the source position is constrained to be no smaller than the pixel of the finest-resolution map used to derive the positions of the flares in the cluster. Systematic errors on the source position are estimated comparing the positions of the 2FAV sources to those of known flaring gamma-ray sources, resulting in a systematic error $r_{sys}=0.1\degree$.

\subsection{Association procedure}\label{subsec:associations}

As for the previous catalog, we provide candidate associations for the catalog sources. We look for counterparts of 2FAV sources in the Third (3FGL), the Second~\citep[2FGL,][]{Nolan_2012} and the First~\citep[1FGL,][]{Abdo_2010} \textit{Fermi} LAT Source catalogs, in the First and the updated AGILE Catalogs \citep[1AGL and 1AGLR,][respectively]{1agl, agile_upvar}, the Third EGRET catalog \citep[3EG,][]{3eg}, the 5$^{th}$ edition of the \textit{Roma-BZCAT} \citep[5BZ,][]{5bz}, as well as in Astronomer's Telegrams (ATels) based on Fermi-LAT observations\footnote{The list of ATels used was current as of 2016-05-30. The most recent version is available at: \href{http://www.asdc.asi.it/feratel/}{http://www.asdc.asi.it/feratel/}}, and among LAT detected gamma-ray bursts (GRBs)\footnote{\href{http://fermi.gsfc.nasa.gov/ssc/observations/types/grbs/lat\_grbs/}{http://fermi.gsfc.nasa.gov/ssc/observations/types/grbs/lat\_grbs/}}. FAVA sources are associated to the closest counterpart found within the search radius $R_{s}$, defined as the sum of the 99\% statistical error on the FAVA source position, plus the systematic uncertainty: $R_{s}=r_{99}+r_{sys}$. This association procedure is based solely on positional coincidence. It does not take into account temporal or spectral properties of the sources. For this reason, its results have to be interpreted only as likely counterparts, rather than firm associations. 

The association procedure is as follows. We search for counterparts of FAVA sources using the \textit{Fermi} LAT catalogs. For each of these catalogs, the search is initially restricted to sources that have less than 1\% probability of being constant on monthly time scales\footnote{Such sources are selected requiring variability indexes greater than 23.21, 41.64, 72.44 in the 1FGL, 2FGL, and 3FGL respectively, resulting in 241, 458 and 647 variable sources in each of these catalogs.}. The search starts with the 3FGL, and uses less-ecent catalogs in case no counterpart is found. We found 352 and 5 2FAV sources associated with variable sources from the 3FGL and 2FGL, respectively. No 2FAV source has been associated to a variable 1FGL source.

If no variable counterpart is found in any of the LAT source catalogs, we search among new LAT sources announced via ATels and in the 5BZ, finding additional counterparts for 16 and 39 sources respectively. For 2FAV sources that are still unassociated, the search among the LAT catalogs is repeated, this time considering all the FGL sources, not only the variable ones. With a much larger sample of possible counterparts, these associations have a higher probability to be spurious. For this reason the name of the LAT catalog counterpart will be flagged with ``*''. With this procedure we find 12 and 1 non-variable counterparts of 2FAV sources in the 3FGL and 2FGL, respectively. Finally the 1AGLR, 1AGL, and 3EG catalogs are used, in this order. We found one counterpart for a 2FAV source in the 1AGL and one in the 3EG.

Finally, associations with LAT detected GRBs are tested. To be associated with a GRB, the 2FAV source must have only a single flaring event coincident in time (within the weekly binning) with the time of the GRB. Moreover, the distance between the source and the GRB must be smaller than the combined error on the GRB localization, plus the source search radius. 14 FAVA sources have been associated with GRBs. In none of these cases alternative counterparts were found in the other catalogs.

\section{The second FAVA catalog}\label{sec:catalog}

\begin{figure}[h!]
\includegraphics[width=1\columnwidth]{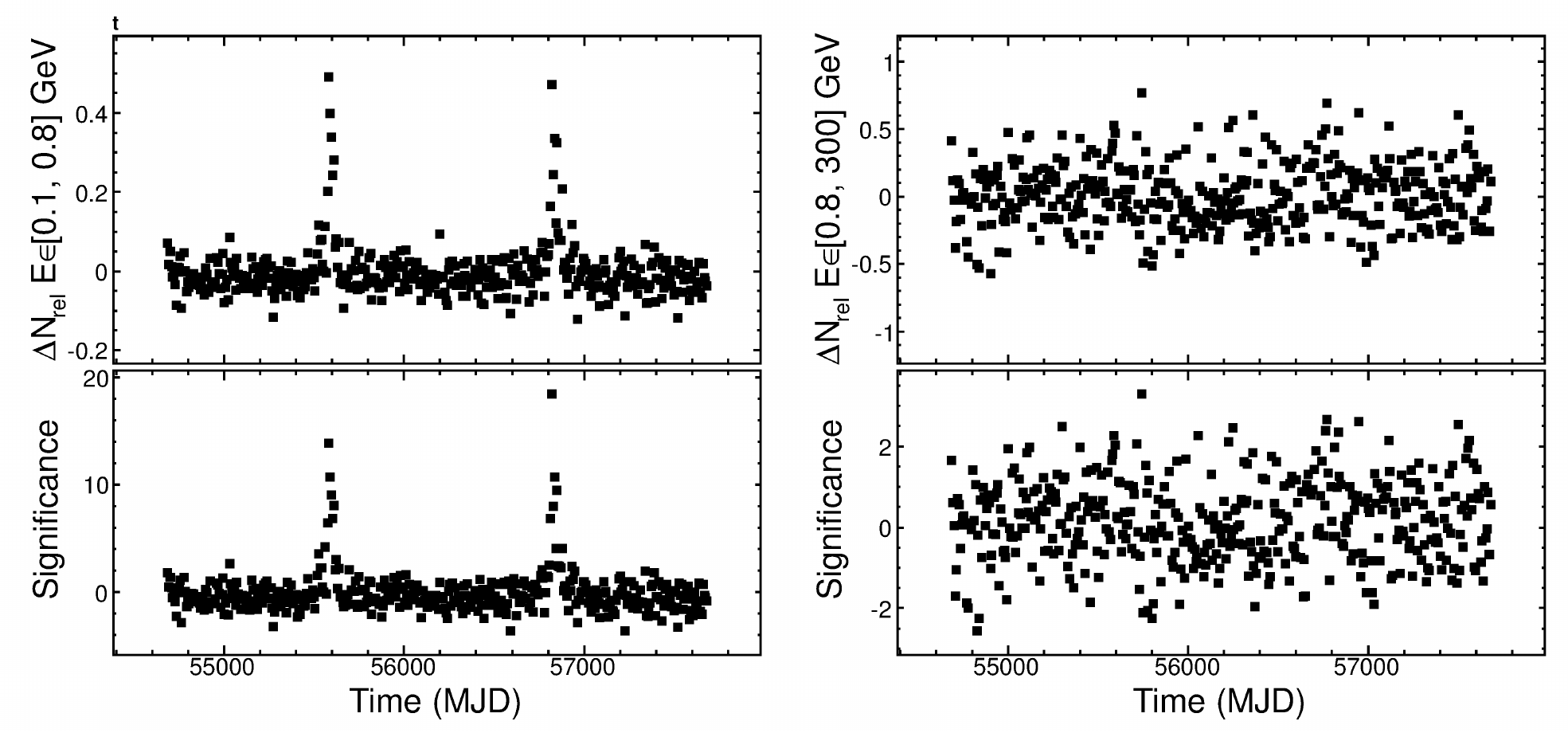}
\caption{Weekly light curves from the direction of the pulsar binary system PSR B1259$-$63/LS 2883 (2FAV J1302$-$63.7). The top panels show the relative variation of counts: $\Delta N_{rel}=(N-N^{exp})/N^{exp}$, where $N$ is the number of observed counts in a given time bin, and $N^{exp}$ is the number of expected counts. The bottom panels show the significance that corresponds to these counts variations. Plots on the left refer to low-energy analysis, while the ones on the right refer to high-energy analysis.}
\label{fig:lc_PSRB125963}
\end{figure}

The second FAVA catalog is presented in Table~\ref{tab:cat_table}. 2FAV sources are named as 2FAV JHHMM+DD.d where HHMM expresses the Right Ascension (J2000) of the source in hours and minutes, and DD.d is the declination in degrees, truncated to the first decimal place. Figure~\ref{fig:catandflares} presents the position of catalog sources and flares. For an all-sky grid of coordinates we produce photometric aperture light curves of the relative flux variations in both energy bands and make them available online\footnote{\href{http://fermi.gsfc.nasa.gov/ssc/data/access/lat/FAVA/CatalogView\_2FAV.php}{http://fermi.gsfc.nasa.gov/ssc/data/access/lat/FAVA/CatalogView\_2FAV.php}}. An example of these light curves can be seen in Figure~\ref{fig:lc_PSRB125963}.

Out of the 518 sources constituting the catalog, 155 have been detected at low energy only. Negative flares are expected from the most variable sources, as bright and frequent flares can push the long-term average flux above the level of the quiescent emission of the source. Negative flares have been recorded for 35 sources, most of them belongings to the Flat Spectrum Radio Quasar (FSRQ) and BL Lacertae (BL Lac) classes, but also from the Crab pulsar wind nebula (2FAV J0534+21.9) and the high-mass X-ray binary system (HMB) LS I+61 303 (2FAV J0240+61.4). 

Within the catalog there are 9 flaring sources localized solely by the photometric FAVA analysis. Each of these sources is associated with a single flaring event. Of the 9 flares constituting these sources, 4 have been detected only at low energies and 5 have a significance greater than $4\sigma$ in both energy bands. For most of these flares, the TS is below the threshold and the 95\% CL contour closes on the border of the map. As the disagreement between the two analyses suggests caution with these sources, they are flagged adding a \lq f\rq{ }to their name, for example 2FAV J2350$-$06.1f.

The results of the association procedure are presented in Table~\ref{tab:catassoc}. For the source classes we follow the classifications used in the 3FGL, unless otherwise noted. Counterparts have been found for 441 sources, from 13 different classes. Of these, 395 have a likely counterpart belonging to the Active Galactic Nucleus (AGN) class, making up $90\%$ of the associated 2FAV sources. Among these, the most represented class is FSRQs. Among the counterparts of 2FAV sources, 16 (12 from the 3FGL, 1 from the 3EG, {1 from 1AGL}, and 2 from ATels) are published as unassociated in their respective catalogs. In the 3FGL, unassociated sources make up roughly 33\% of the whole list, yet they only represent $\simeq3\%$ of the counterparts of 2FAV sources. Unassociated 3FGL sources, regardless of class, are less likely to be flaring gamma-ray emitters, at least on the weekly time scales.

Among the FAVA sources that have non-blazar counterparts we find three Narrow-line Seyfert 1 galaxies: 2FAV J0948+00.3, 2FAV J0849+51.1, and 2FAV J0324+34.2 associated with PMN J0948+0022~\citep{DAmmando_2015}, SBS 0846+513~\citep{DAmmando_2012}, and 1H 0323+342 \citep{1H_0323+342} respectively, and three radio galaxies: 2FAV J0419+38.2, 2FAV J0319+41.5, and 2FAV J0433+05.4 associated with 3C 111~\citep{Grandi_2012}, NGC 1275 \citep{NGC1275}, and 3C 120 \citep{3c120_6yr} respectively. The 14 GRBs included in the 2FAV catalog all have energy fluence, computed in the energy range 0.1$-$100 GeV, in excess of $1.4\times10^{-5}$ erg~$\text{s}^{-1}$. They are among the brightest ones detected by the LAT~\citep{lat_grb_first_cat}.

Five 2FAV sources are associated with LAT-detected novae~\citep{latnovae, latnovae2}: 2FAV J2102+45.7 \citep[V407 Cyg,][]{V407Cyg}, 2FAV J0639+05.8 \citep[V959 Mon 2012,][]{nova_mon_2012}, 2FAV J1751$-$32.5 \citep[V1324 Sco 2012,][]{nova_sco_2012}, 2FAV J2023+20.7 \citep[V339 Del 2013,][]{nova_delphini}, and 2FAV J1354$-$59.1 \citep[V1369 Cen 2013,][]{nova_centauri}. Compared to the list of LAT-observed novae\footnote{\href{http://asd.gsfc.nasa.gov/Koji.Mukai/novae/novae.html}{http://asd.gsfc.nasa.gov/Koji.Mukai/novae/novae.html}}, we found no other nova that would be expected to be included in the catalog, at the high significance threshold we are using. 

Other Galactic 2FAV sources include the Crab nebula \citep[2FAV J0534+21.9,][]{Abdo_2011, crab_agile_2011, Buehler_2012, Mayer_2013}, and three gamma-ray emitting binaries: 2FAV J2032+40.9 associated with Cyg X$-$3 \citep{Abdo_2009, cygx3_agile}, 2FAV J0240+61.4 associated with LS I+61 303~\citep{LSIofficial}, and 2FAV J1302$-$63.7 associated with the pulsar/Be-star binary system PSR B1259$-$63/LS 2883 \citep{PSRB125963_first, Caliandro_2015}. The gamma-ray binaries LS 5039~\citep{LS5039} and 1FGL J1018.6$-$5856~\citep{1FGLJ1018.6-5856} are not part of the 2FAV catalog (they were not in 1FAV either).

Two FAVA sources are found to be positionally coincident with LAT-detected pulsars when the search for counterparts has been repeated to include non-variable 3FGL sources. 2FAV J1023+00.6 is associated with the millisecond pulsar binary PSR J1023+0038, whose gamma-ray flux increased in June/July 2013~\citep{PSR_J1023+0038_LAT}, possibly due to the propeller effect~\citep{prop1, prop2} or the development of an accretion disk~\citep{Takata_2014}. The other pulsar which shows gamma-ray flux variability, PSR J2021+4026~\citep{PSR_J2021+4026}, is not included in this catalog. Due to the reduced sensitivity of FAVA along the Galactic plane (PSR J2021+4026 is at $b\simeq2.1\degree$) the $\approx20\%$ flux drop observed for this source around MJD 55850 has a significance of $\approx3.8\sigma$\footnote{\href{http://fermi.gsfc.nasa.gov/ssc/data/access/lat/FAVA/LightCurve.php?ra=305.386\&dec=40.448}{http://fermi.gsfc.nasa.gov/ssc/data/access/lat/FAVA/LightCurve.php?ra=305.386\&dec=40.448}}, just below the threshold used in this analysis.
2FAV J1824$-$13.0 is positionally coincident with the pulsar PSR J1826$-$1256, for which no evidence of flaring activity in gamma rays has been found to date. Two low-energy flares have been associated with 2FAV J1824$-$13.0. However, only in one case the pulsar was the only source contained inside the 95\% contour of the TS map. Moreover, during the weekly time bin when the flare occurred, the LAT was performing a long target of opportunity (ToO) observation of V5668 Srg 2015~\citep[see][]{latnovae2}, pointing only 2.3 degrees from the pulsar. As the association of the flaring event with PSR J1826$-$1256 can not be established firmly, 2FAV J1824$-$13.0 is listed as unassociated in this catalog. For future occurrences, prompt follow-up observations to determine eventual correlated variability at other wavelengths will be crucial to establish the source of the outburst. 

In the 2FAV catalog there are 6 sources that have counterparts from the 2FGL. Three of these counterparts are sources that have been \lq lost\rq\ in the 3FGL \citep[see][sec. 4.2]{Acero_2015}. In the other cases the 2FGL counterpart of the 2FAV source is also present in the 3FGL. The former 2FAV sources are not directly associated with the 3FGL counterpart either because the source has no longer been found variable in the 3FGL, or because the 3FGL position is outside of the search radius. These details are presented in Table~\ref{tab:lostfgl}.

To evaluate the improvements with respect to the 1FAV, we compare the lists of flares detected by the two versions of FAVA in the same time span, the first 47 months of \textit{Fermi} observations. For this common time interval, the number of detected flares has increased by $\sim43\%$ (from 1419 in the 1FAV to 2025 in the 2FAV), despite the lower significance threshold used in the 1FAV ($5.5\sigma$). The increase in the number of detected flares is mostly due to the addition of the likelihood analysis and to the inclusion of $4\sigma$ flares found simultaneously in both energy bands. The independence of the two energy bands of the 2FAV analysis makes it possible to combine two weak ($4\sigma$), spatially coincident, and simultaneous detections in the two energy bands into a more significant one. The effect of the new Pass 8 data release on the photometric analysis is estimated by comparing the significances and the localization accuracy (by means of the source-flare distance) of the flares in the two catalogs. The harmonic mean of the source-flare distance has improved from $0\fdg42$ in the 1FAV to $0\fdg34$ in the 2FAV. The high-energy significances of 2FAV flares are also $\sim6\%$ higher than those measured for the same flares in the 1FAV. In the low-energy band, no appreciable improvement of the photometric significance has been found.
Considering the sources in the two FAVA catalogs, for 15 of the 215 1FAV sources, no 2FAV counterpart is found within the combined search radius. The lower significance threshold used to construct the 1FAV catalog accounts for 10 of these cases. The flares associated with the remaining 5 1FAV sources either have a 1FAV significance which is just above $6\sigma$ in 1FAV, that could have gone below threshold in the 2FAV, or are spatially coincident with some 2FAV sources (this can happen even if the 1FAV source to which these flares were assigned is not positionally compatible with the 2FAV one). In the second case the 1FAV flares are not lost in the 2FAV and the discrepancies between the two catalogs are due to a different assignments in the clustering of flares that moved the 2FAV source away from the 1FAV position.

For 77 2FAV sources, no counterpart has been found in the catalogs used for the association search. The spectral properties of the flares of these potentially new sources are presented in Section~\ref{subsubsec:nnblz_spec}. The great majority of these sources are located outside of the Galactic plane, with only 2 sources with $|b|<5\degree$: 2FAV J1259$-$65.4 and 2FAV J2010+35.7. A dedicated search for counterparts of these low-latitude sources suggests as a reasonable counterpart candidate for 2FAV J2010+35.7 the compact radio source B2 2008+35~\citep[VERA J20089+3543,][]{VERA22}, 4.8 arcmin away from the 2FAV source. No plausible counterpart candidate has been found for 2FAV J1259$-$65.4.

Our analysis covers a much wider time span than the one used to construct the 3FGL. This could, at least partially, explain the relatively large number of potentially new sources found. Twenty-one of these unassociated sources flared solely during the 3FGL period, while for 49 of them the first measured flare happened after the 3FGL time span. The great majority of the 21 sources that flared only during the 3FGL period flared only once; in only one case (2FAV J0905+01.3), two flares were detected.

\section{Flare spectra}\label{subsec:flarespec}

In this Section we study the spectral properties of the flares associated to the 2FAV sources. The high-energy spectra are not corrected to account for absorption on the Extragalactic Background Light (EBL). The farthest associated 2FAV source is 2FAV J0539$-$28.8 (FSRQ PKS 0537$-$286) at a redshift $z=3.1$. For this redshift, EBL absorption affects the shape of spectra above $\approx6$ GeV~\citep{EBL}. Spectral measurements in the low-energy band are therefore unaffected. Since the high-energy FAVA band reaches down to 800 MeV, the measure of the spectral slope in this energy band is still dominated by the lowest energies. Even for the farthest source, the EBL is expected to have little effect, and will be neglected.

\subsection{Blazars}\label{subsubsec:blz_spec}

\begin{figure}[h!]
    \begin{minipage}[b]{0.5\columnwidth}
    \centering
    \includegraphics[width=\columnwidth]{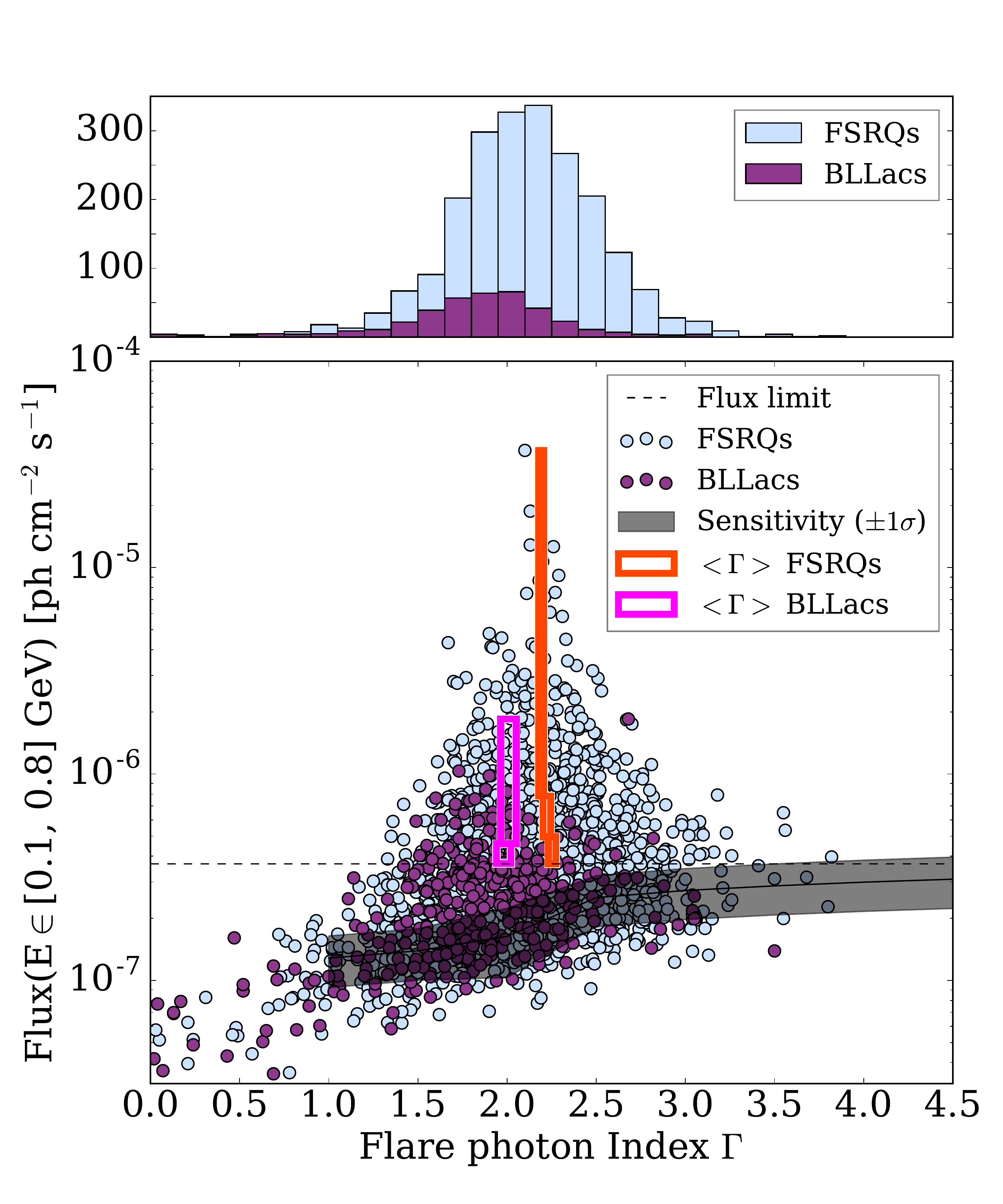}
    \end{minipage}
    \hfill
    \begin{minipage}[b]{0.5\columnwidth}
    \centering
    \includegraphics[width=\columnwidth]{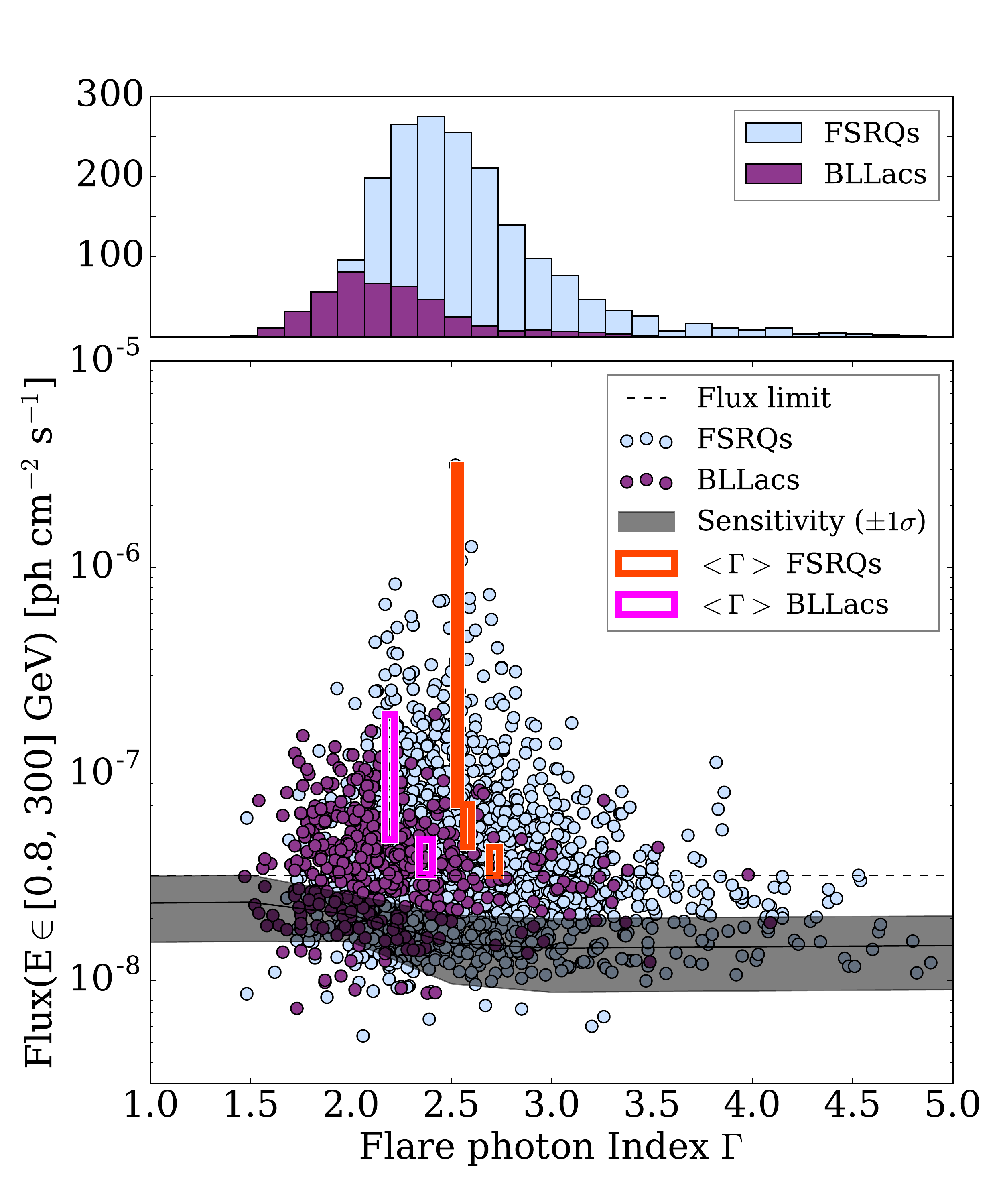}
    \end{minipage}
    \caption{Parameters of the low-energy (left) and high-energy (right) spectra for flares of 2FAV sources associated with FSRQs and BL Lacs. Individual flares are plotted in shades of blue. The gray band represents the average sensitivity (not accounting for the likelihood follow-up) computed at the position of the sources (solid black line), plus or minus one standard deviation. The dotted line represents the flux limit, chosen as the average sensitivity plus one standard deviation for $\Gamma=3.5$ and $\Gamma=1.5$ at low and high energies respectively, see text for details. Magenta and violet boxes show the mean photon index for different flux bins above the flux threshold (black dotted line) for FSRQs and BL Lacs respectively. The size of the boxes indicates the width of the flux bin (in the $y$-direction) and the error on the mean photon index ($x$-direction). The top panels show the distribution of the photon index of the flares in the two respective energy bands.}
    \label{fig:flarespec_blazar}
\end{figure}

To account for the instrumental flux limit, we compute photometric sensitivity maps such as those on Fig.~\ref{fig:sensmaps} for a range of flare photon indexes. For each value of the flare photon index, the sensitivities at the positions of the 2FAV blazars are read from the corresponding map. The average sensitivity and the spread of its distribution are presented as a solid line and gray band in Figure~\ref{fig:flarespec_blazar}. The flux limits in the low- and high-energy bands are derived as the average sensitivity plus one standard deviation for the maps corresponding to $\Gamma=3.5$ and $\Gamma=1.5$ respectively. These values of the flare photon index give conservative estimates of the sensitivity in the two energy bands (see the dotted line in Figure~\ref{fig:flarespec_blazar}). The resulting flux limits are $F^{sens}_{LE}=3.67\times10^{-7}\text{cm}^{-2}\text{s}^{-1}$ in the low-energy band, and $F^{sens}_{HE}=3.24\times10^{-8}\text{cm}^{-2}\text{s}^{-1}$ at high energies. Note that our sensitivity calculation does not account for the likelihood analysis, which is independent to the photometric one. As a consequence, these flux limits are a conservative estimate of the sensitivity of the entire 2FAV analysis.

Figure~\ref{fig:flarespec_blazar} shows the spectral parameters of low- and high-energy flares from 2FAV sources associated with BL Lacs and FSRQs. We see that BL Lac flares are on average harder than the ones from FSRQs, a difference already observed in the time-averaged spectra of these sources. Above the sensitivity threshold, the mean and standard deviation of the photon index distributions for FSRQs and BL Lac flares are respectively: $2.19\pm0.33$ and $1.97\pm0.27$ at low energies, and $2.50\pm0.36$ and $2.14\pm0.32$ at high energies. These values are in agreement with the average photon indexes for the entire BL Lac and FSRQ populations found in the Third \textit{Fermi} LAT AGN catalog \citep[3LAC,][]{3lac}. The energy bands of the 2FAV differ from the ones used in the 3LAC and 3FGL catalogs. As a consequence, a source-wise comparison between the photon indexes of the flares and those corresponding to the time-integrated emission is not possible.

We measure a variation of photon index with flux level. The samples of flares above the sensitivity limits are divided in equally populated flux bins (three for FSRQs and two for BL Lacs), and the mean flare photon index is computed for each bin. The difference between the average photon index in the lowest- and highest-flux bin has a significance of $2.4\sigma$ for FSRQs and $0.4\sigma$ for BL Lacs at low energies, and of $6.2\sigma$ for FSRQs and $4.3\sigma$ for BL Lacs at high energies. The \lq harder when brighter\rq{ }behavior, often observed for individual sources \citep[see, e.g,][]{mkn501magic, pks2155hess, 3c4543fermi}, is apparent for the entire samples of flares with the exception of low-energy BL Lacs flares. In this case, the flare sample is dominated by three objects\footnote{2FAV J2202+42.2 (BL Lacertae) , 2FAV J0238+16.6 (AO 0235+164), and 2FAV J0428-37.9 (PKS 0426$-$380)} which together account for $\approx60\%$ of the 82 flares in the sample. Of these three sources, only for 2FAV J0238+16.6 do the spectra of its flares show a significant hardening for higher fluxes.	

According to the position of the synchrotron peak in the Spectral Energy Distribution (SED), blazars can be classified into low-synchrotron-peaked (LSP, for sources with $\nu_{peak}<10^{14}$~Hz), intermediate-synchrotron-peaked (ISP, $10^{14}\text{Hz}<\nu_{peak}<10^{15}$~Hz), and high-synchrotron-peaked (HSP, $\nu_{peak}>10^{15}$~Hz). Using the 3LAC we find SED classifications for 349 2FAV sources: 295 LSP, 46 ISP, and 19 HSP. We use the method described above (with 3 flux bins for LSP and 2 for ISP and HSP) to measure the \lq harder when brighter\rq{ }behavior for the different SED classes. The results are summarized in Table~\ref{tab:harderwhenbrighter_SED}. A significant effect is measured only for high-energy flares of LSP blazars.

Above the sensitivity threshold, no flare with photon index harder than $\sim$1.5 is detected. This is consistent with what is already known about the time-averaged spectra of blazars \citep[see e.g.][]{AGNinGamma}. For photon indexes $<2$ the SED is rising. In leptonic models this implies that the electron-photon interaction is likely to happen in the Thomson regime. Under the assumption of isotropic electron and seed photon fields, we have $\Gamma=(p+1)/2$~\citep{radiativeProc}, where $p$ is the power-law index of the electron spectrum $dN/dE$. The observed limit $\Gamma\gtrsim1.5$ implies $p\gtrsim2$. This is compatible with both shock and low-magnetization magnetic acceleration of particles \citep[for reviews, see][]{diffShocks, blandford_shocks, relshocks, magreconn}.

\subsection{Non-blazar sources}\label{subsubsec:nnblz_spec}

\begin{figure}[h!]
\centering
\includegraphics[width=1\columnwidth]{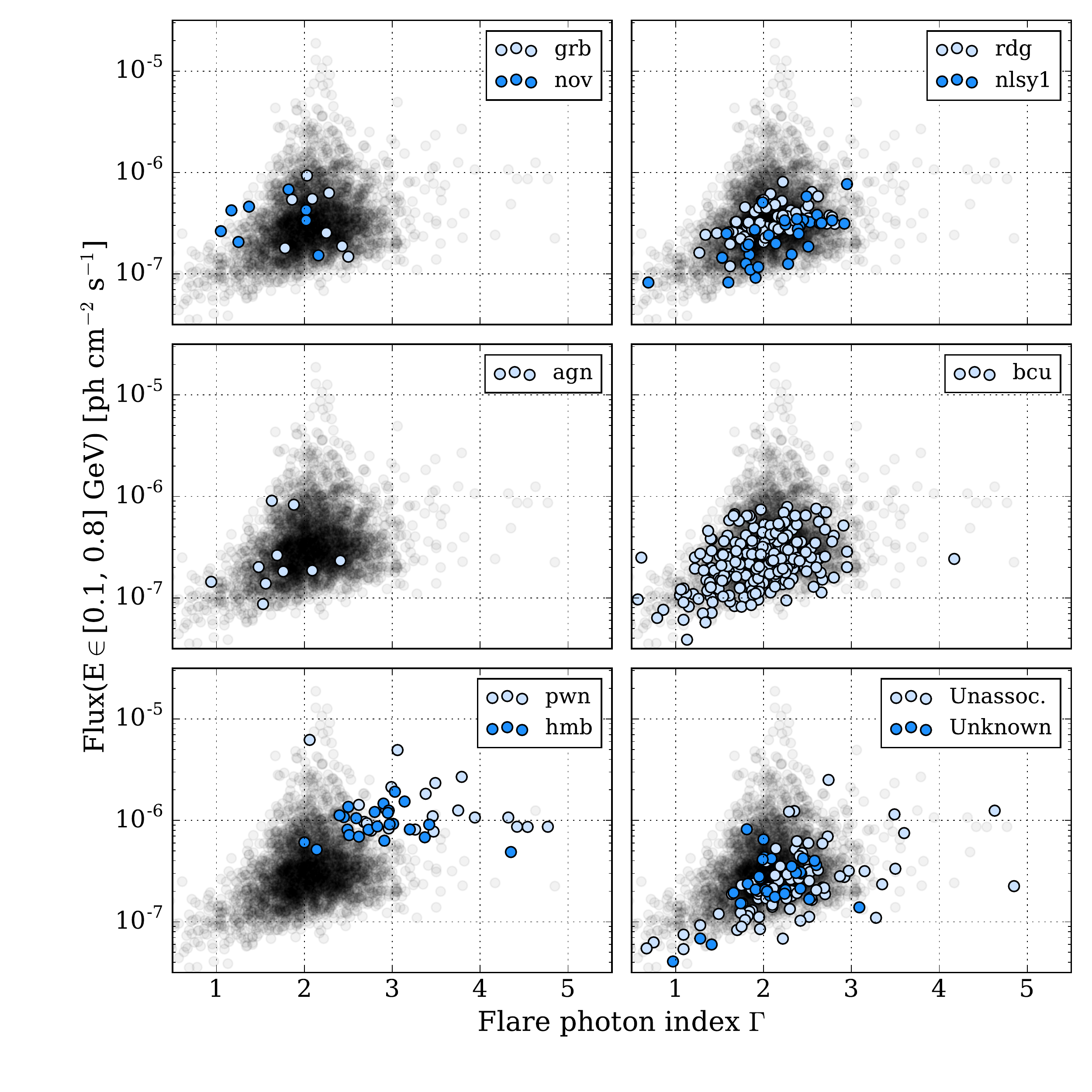}
\caption{Parameters of the low-energy spectra for the flares of the 2FAV sources (for BL Lacs and FSRQs, see Fig.~\ref{fig:flarespec_blazar}). The flares belonging to each source class are plotted in colors. In gray is plotted the entire sample of 2FAV flares detected in this energy band. Sources tagged with \lq Unknown\rq~(last panel) are 2FAV sources whose counterpart has not been associated in the respective catalogs.}
\label{fig:flarespec_all_le}
\end{figure}

\begin{figure}[h!]
\centering
\includegraphics[width=1\columnwidth]{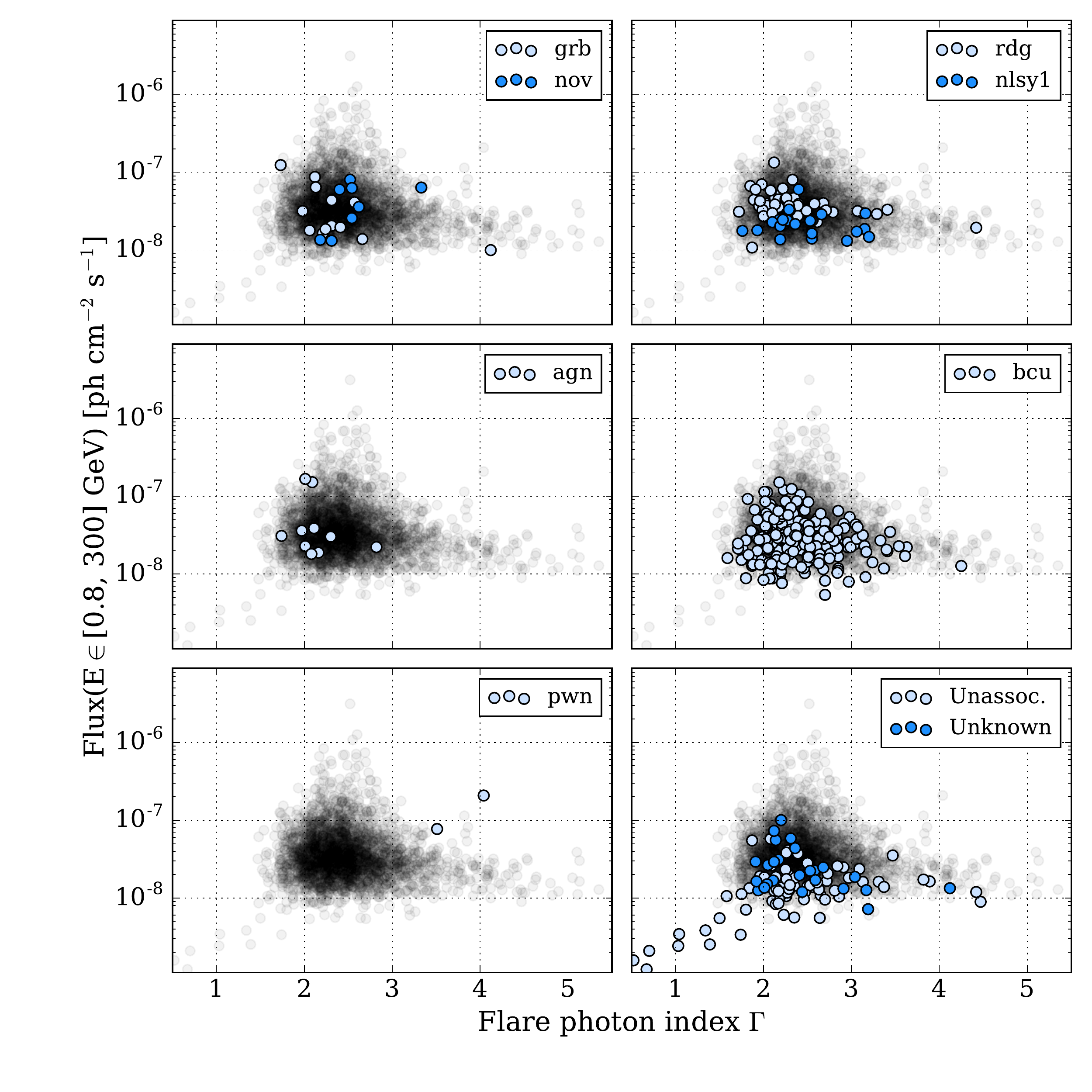}
\caption{Parameters of the high-energy spectra for the flares of the 2FAV sources (for BL Lacs and FSRQs, see Fig.~\ref{fig:flarespec_blazar}). See Fig.~\ref{fig:flarespec_all_le} for a description of the legends.}
\label{fig:flarespec_all_he}
\end{figure}

The distribution of spectral parameters for the flares of all the other 2FAV sources is presented in Figures~\ref{fig:flarespec_all_le} and \ref{fig:flarespec_all_he}. Interestingly, low-energy flares associated with the Crab Nebula and the three binary systems Cyg X$-$3, LS I+61 303, and PSR B1259$-$63/LS 2883 populate a different region of the parameter space than the one occupied by the rest of the sources (see lower left panel of Fig.~\ref{fig:flarespec_all_le}). This could hint at differences in the emission mechanism responsible for the flares. In the case of the Crab nebula for example, magnetic reconnection is preferred over shock acceleration \citep{buehler_blandford_crabrev}.

In both energy bands, the flares from sources associated with active galaxies of non-blazar type (agn) are harder than those from blazars. The medians of the photon index distributions for agn flares are 1.7 and 2.1, at low and high energy respectively. The medians of the photon index for the entire blazar flare sample are 2.1 and 2.4, in the low- and high-energy bands, respectively. Flares from sources associated with radio galaxies (rdg) have median low-energy photon index of 2.0, similar to that of blazars, but are harder in the high-energy band (with a median of 2.2). Flares associated with Narrow-line Seyfert 1 galaxies (nlsy1) also appear less curved: with a median photon index of 2.2 they are softer than those of blazars in the low-energy band, but have compatible median (2.4) at high energies.

As visible in the bottom-left panel of Figure~\ref{fig:flarespec_all_he}, some high-energy flares of unassociated sources are faint and hard. Their flux in the 0.8$-$300 GeV band is less than $1\times10^{-8}$ $\text{ph}~\text{cm}^{-2}\text{s}^{-1}$ and their photon index is smaller than 2. Most of these flares are included in the catalog because they have been detected with TS $>$ 18 simultaneously in the two energy bands. The test statistic of the high-energy detection is TS $\sim$ 20 on average. The statistical errors of the spectral measurement are therefore large. The high-energy photon indexes of these flares are compatible with $\Gamma = 1.5$ within less than 1.5 sigma in all cases.

\section{Outlook and conclusions}

FAVA analysis has been upgraded with the addition of an automated likelihood follow-up analysis for flares above $4\sigma$. This has improved the flare localization accuracy by a factor of 3 and 8, at low and high energy respectively. Making use of the Pass 8 data and IRFs, FAVA has been applied to the search for transient and variable emission on weekly time scales over the first 7.4 years of the \textit{Fermi}-LAT mission. 

A total of 4547 flares has been detected with significance above $6\sigma$. Clustering the positions of these flares, 518 flaring gamma-ray sources have been found. For each of these sources the catalog provides possible counterparts and a detailed list of the gamma-ray flares associated to the source. These results, as well as photometric aperture light curves for an all-sky grid of coordinates, are made available online\footnote{\href{http://fermi.gsfc.nasa.gov/ssc/data/access/lat/FAVA/CatalogView\_2FAV.php}{http://fermi.gsfc.nasa.gov/ssc/data/access/lat/FAVA/CatalogView\_2FAV.php}}. For 77 2FAV sources, no counterparts have been found in the catalogs used for the association procedure.

A harder-when-brighter behavior has been observed in the spectra of the collective sample of FSRQ flares, and for the high-energy flares from BL Lacs. No flare with a spectrum significantly harder than $\Gamma\leq1.5$ has been detected. Under the assumption of a leptonic model with isotropic particle distribution, this implies that the number energy spectrum of the freshly accelerated electrons is never harder than $\sim2$. This is compatible with shock acceleration models and magnetic reconnection scenarios.

To maintain compatibility with the 1FAV, we applied FAVA on weekly time bins. However, FAVA can be used to monitor short-term flares down to time scales of a few hours, as well as longer time scales of a few months, covering variability at time scales that are not currently being monitored by the LAT Collaboration. Future plans include running FAVA for shorter time scales and making the results publicly available. With its all-sky view, and fast and robust analysis running on-line, FAVA could provide rapid alerts to the community. The online tool, in particular, will allow users who study variable sources at other wavelengths to quickly search the \textit{Fermi} data for correlated variability.
\acknowledgments
The \textit{Fermi} LAT Collaboration acknowledges generous ongoing support from a number of agencies and institutes that have supported both the development and the operation of the LAT as well as scientific data analysis. These include the National Aeronautics and Space Administration and the Department of Energy in the United States, the Commissariat \`a l'Energie Atomique and the Centre National de la Recherche Scientifique / Institut National de Physique Nucl\'eaire et de Physique des Particules in France, the Agenzia Spaziale Italiana and the Istituto Nazionale di Fisica Nucleare in Italy, the Ministry of Education, Culture, Sports, Science and Technology (MEXT), High Energy Accelerator Research Organization (KEK) and Japan Aerospace Exploration Agency (JAXA) in Japan, and the K.~A.~Wallenberg Foundation, the Swedish Research Council and the Swedish National Space Board in Sweden. 
\\
Additional support for science analysis during the operations phase is gratefully acknowledged from the Istituto Nazionale di Astrofisica in Italy and the Centre National d'\'Etudes Spatiales in France.

\newpage
\thispagestyle{empty}
\bibliographystyle{apj}
\bibliography{2fav}


\begin{table}[h]
\begin{center}
	\begin{tabular}{cccc|c}
        Energy band & Likelihood & FAVA (positive flares) & FAVA (negative flares)  &  total \\
        \midrule
        High Energy & 1748        	& 18            & 228     & 1994 \\
		Low Energy  & 1156			& 57            & 579     & 1792 \\
        Combined    & 524         	& 77            & 160     & 761 \\
        \hline
	    total       & 3428         	& 152			& 967     & 4547 \\
    \end{tabular}
    \caption{Number of catalog flares in the different cut categories. As a single flare can satisfy more than one cut, to construct this table the cuts are made mutually exclusive in order to not count the same flare more than once. A flare is not counted as passing cuts on its photometric FAVA properties if it already satisfies some cut on its likelihood analysis results. For each analysis flare properties in different energy bands are considered in this order: high energy, low energy, and combined energy range. The last column shows the total number of flares detected in the different energy bands. The bottom row shows the total number of flares passing cuts for TS maps, and for positive and negative photometric FAVA analysis respectively.\label{tab:flarecuttable}}
\end{center}
\end{table}


\begin{deluxetable}{ l c c c c c c c c c l l c c }
\tabletypesize{\scriptsize}
\tabcolsep=0.1cm
\rotate
\tablecaption{List of 2FAV sources. The first column shows the FAVA identification number (ID). The following columns show the position of the source in Equatorial (J2000) and Galactic coordinates, and the statistical error on the source position at 95\% confidence level $r_{95}$. The total number of flares ($\text{N}^{f}$) detected for that source is shown, together with the number of high-energy flares ($\text{N}^{f}_{he}$), and the number of negative flares ($\text{N}^{f}_{neg}$). The last four columns show the search radius used in the association procedure ($\text{R}_{s}$, see text for details), the name of the counterpart found in the gamma-ray catalogs used for the association (Gamma assoc.), the name of the counterpart at other wavelengths as published in the considered catalogs (Assoc.), the class of the counterpart (class), and the ATel numbers for \textit{Fermi}-LAT flare alerts eventually associated with the source. Sources localized solely by the photometric FAVA analysis are marked adding a \lq f\rq~ to the ID number. \label{tab:cat_table}}
\tablewidth{0pt}
\tablehead{
\colhead{2FAV ID} &
\colhead{RA [$\degree$]} &
\colhead{Dec [$\degree$]} &
\colhead{\textit{l} [$\degree$]} &
\colhead{\textit{b} [$\degree$]} &
\colhead{$r_{95} [\degree$]} &
\colhead{$\text{N}^{f}$} &
\colhead{$\text{N}^{f}_{he}$} &
\colhead{$\text{N}^{f}_{neg}$} & 
\colhead{$\text{R}_{s}$ [$\degree$]} &
\colhead{Gamma assoc.} &
\colhead{Assoc.} &
\colhead{class} &
\colhead{ATel}}
\startdata
J0001+20.6 & 0.39 & 20.65 & 107.50 & $-$40.70 & 0.87 & 1 & 0 & 0 & 1.18 & 3FGL J0001.4+2120 & TXS 2358+209 & fsrq & ...\\
J0002+75.0 & 0.71 & 75.03 & 119.70 & 12.50 & 0.14 & 1 & 1 & 0 & 0.27 & ... & ... & ... & ...\\
J0017$-$05.1 & 4.48 & $-$5.15 & 101.50 & $-$66.60 & 0.13 & 6 & 1 & 0 & 0.27 & 3FGL J0017.6$-$0512 & PMN J0017$-$0512 & fsrq & ...\\
J0019+73.4 & 4.97 & 73.48 & 120.70 & 10.70 & 0.05 & 5 & 4 & 0 & 0.16 & ... & 5BZQ J0019+7327 & fsrq & ...\\
J0025$-$48.1 & 6.42 & $-$48.10 & 314.70 & $-$68.40 & 0.08 & 2 & 2 & 0 & 0.20 & 3FGL J0026.2$-$4812* & ... & ... & ...\\
J0028+06.9 & 7.21 & 6.93 & 113.00 & $-$55.50 & 0.84 & 1 & 0 & 0 & 1.14 & ... & 5BZQ J0029+0554 & fsrq & ...\\
J0030$-$02.2 & 7.66 & $-$2.22 & 110.80 & $-$64.60 & 0.05 & 11 & 7 & 0 & 0.16 & 3FGL J0030.7$-$0209 & PKS B0027$-$024 & bcu & ...\\
J0032$-$55.2 & 8.12 & $-$55.29 & 308.60 & $-$61.60 & 0.08 & 2 & 2 & 0 & 0.19 & 3FGL J0032.3$-$5522 & ... & ... & ...\\
\enddata
\end{deluxetable}


\begin{deluxetable}{ l   c  r }
\tablecaption{Association classes for the 2FAV sources. The labels of the various classes are the same as the ones used in the 3FGL.\label{tab:catassoc}}
\tablewidth{0pt}
\tablehead{
\colhead{Source class} &
\colhead{Designator} &
\colhead{Number of associated sources}}
\startdata
Pulsar wind nebula & pwn & 1 \\
Pulsar & psr & 1 \\
Non-blazar active galaxy & agn & 2 \\
High-mass binary & hmb & 3 \\
Radio galaxy & rdg & 3 \\
Narrow line Seyfert 1 & nlsy1 & 3 \\
Nova & nov & 5 \\
Gamma-ray burst & grb & 14 \\
Blazar candidate of uncertain type & bcu & 67 \\
BL Lac & bll & 73\\
FSRQ & fsrq & 253 \\
\hline
Unassociated counterpart & Unknown & 16 \\
Unassociated & \nodata & 77 \\
\enddata
\end{deluxetable}


\begin{center}
\begin{deluxetable}{lcccccccc}
\tabletypesize{\scriptsize}
\tabcolsep=0.1cm
\rotate
\tablecaption{List of 2FAV sources that have been associated with 2FGL sources and the possible 3FGL counterpart of the associations. The first two columns show the 2FAV name and the search radius used to search for counterparts. The third, fourth, and fifth columns show the name of the 2FGL counterpart, its distance from the 2FAV source, and its variability index. The last three columns show the same information for the candidate 3FGL source to which the 2FGL counterpart is associated. The last rows of the table refer to sources that have been associated to 2FGL sources that are \lq lost\rq\ in the 3FGL. We note that to be considered variable at the 99\% confidence level, a source must have a variability index greater than 41.64 and 72.44 in the 2FGL and 3FGL respectively~\citep{Nolan_2012, Acero_2015}.\label{tab:lostfgl}}
\tablewidth{0pt}
\tablehead{
\colhead{2FAV ID} &
\colhead{$R_{s}~[\degree]$} &
\colhead{2FGL assoc.} &
\colhead{2FGL dist.~$[\degree]$} &
\colhead{2FGL var. index} &
\colhead{3FGL assoc} &
\colhead{3FGL dist.~[$\degree$]} & 
\colhead{3FGL var. index}}
\startdata
J0709+22.7 & 0.21 & J0709.0+2236  & 0.11 & 45.63 & J0708.9+2239  & 0.07 & 56.60\\
J2022+76.2 & 0.19 & J2022.5+7614  & 0.03 & 57.89 & J2022.5+7612  & 0.02 & 66.59\\
J0059$-$56.8 & 0.43 & J0059.7$-$5700  & 0.13 & 43.01 & J0059.1$-$5701  & 0.15 & 64.37\\
\midrule
J0910$-$50.8 & 0.19 & J0910.4$-$5050  & 0.07 & 59.12 & \nodata & \nodata & \nodata\\
J1023+00.6 & 0.18 & J1023.6+0040  & 0.02 & 15.32 & \nodata & \nodata & \nodata\\
J2032+40.9 & 0.30 & J2032.1+4049  & 0.16 & 121.21 & \nodata & \nodata & \nodata\\
\enddata
\end{deluxetable}
\end{center}

\begin{table}[h]
\begin{center}
	\begin{tabular}{c|cc|cc}
				 & \multicolumn{2}{|c|}{$\text{E}\in [0.1, 0.8]~\text{GeV}$} & \multicolumn{2}{|c}{$\text{E}\in[0.8, 300]~\text{GeV}$} \\
        SED class & Signif. ($\sigma$) & $N_{flares}(N_{sources})$ & Signif. ($\sigma$) & $N_{flares}(N_{sources})$ \\
        \midrule
        LSP 	& 1.6       	& 901 (107)		& 5.4     & 1159 (149) \\
        ISP  	& 0.5			& 65 (7)    	& 0.8     & 28 (10) \\
        HSP   	& 0.2         	& 5 (4)     	& 2.3     & 91 (15) \\
    \end{tabular}
    \caption{Significance of the spectral hardening measured for the highest and lowest flux bin for LSP, ISP, and HSP blazars. The second and third columns refer to flare spectra measured in the low-energy band, while the last two columns to the high-energy band. In the third and fifth column are the numbers of flares (and sources) that are above the flux limit and contribute to the result. See text for details.\label{tab:harderwhenbrighter_SED}}
\end{center}
\end{table}

\end{document}